\documentclass[useAMS,usenatbib]{mn2e}
\usepackage{graphicx,ifthen,dcolumn,url}
\newcommand{\figdir}{.}

\def\aatab#1#2#3#4#5#6{\ifthenelse{\equal{*}{#1}}
{\begin{table*}\caption[]{\label{#2} #3}
  \begin{flushleft}
    \begin{tabular}{#4}
      \hline\noalign{\smallskip} #5
      \noalign{\smallskip} \hline \noalign{\smallskip} #6
      \noalign{\smallskip} \hline
    \end{tabular}
  \end{flushleft}
\end{table*}}
{\begin{table}\caption[]{\label{#1} #2}
  \begin{flushleft}
    \begin{tabular}{#3}
      \hline\noalign{\smallskip} #4
      \noalign{\smallskip} \hline \noalign{\smallskip} #5
      \noalign{\smallskip} \hline
    \end{tabular}
  \end{flushleft}
\end{table}}}

\newcommand{\be}[1]{\begin{equation}\label{#1}}
\newcommand{\ee}{\end{equation}}
\newcommand{\bea}[1]{\begin{eqnarray}\label{#1}}
\newcommand{\eea}{\end{eqnarray}}
\newcommand{\lwig}{{\leavevmode\kern0.3em\raise.3ex\hbox{$<$}
                    \kern-0.8em\lower.7ex \hbox{$\sim$}\kern0.3em}}
\newcommand{\dd} [2]{{{{\rm d}{#1}\over{\rm d}{#2}}}}

\renewcommand{\dd}{{\rm d}}

\newlength{\pwidth}
\newcommand{\ppmm}[2]{\lower-.7ex\hbox{\scriptsize$#1$}\settowidth{\pwidth}%
            {\scriptsize$#2$}\kern-\pwidth\lower.5ex\hbox{\scriptsize$#2$}}
\def\note #1]{{\bf #1]}}
\def\nc{n_{\rm c}}
\def\epsilonc{\epsilon_{\rm c}}
\def\Msun{\, {\rm M}_\odot}
\def\Rsun{\, {\rm R}_\odot}
\def\CMrgb{{\cal M}_{\rm RGB}}
\def\CMrc{{\cal M}_{\rm RC}}
\def\Rs{R_{\rm s}}
\def\CI{{\cal I}}
\def\CJ{{\cal J}}

\begin{document}
    \title[On the asymptotic phase in red-giant stars]
          {On the asymptotic acoustic-mode phase in red-giant stars
and its dependence on evolutionary state}
    \author[J. Christensen-Dalsgaard et al.]{J{\o}rgen Christensen-Dalsgaard$^{1}$\thanks{E-mail: jcd@phys.au.dk},
     Victor Silva Aguirre$^{1}$,
     Yvonne Elsworth$^{2,1}$, \newauthor
     Saskia Hekker$^{3,1}$\\
    $^{1}$Stellar Astrophysics Centre, Department of Physics and Astronomy,
          Aarhus University, DK--8000 Aarhus C, Denmark,\\
    $^{2}$School of Physics \& Astronomy, University of Birmingham,
          Edgbaston Park Road, West Midlands, Birmingham B15 2TT, UK, \\
    $^{3}$Max-Planck-Institut f\"ur Sonnensystemforschung, 
          Justus-von-Liebig-Weg 3, 37077 G\"ottingen, Germany\\
}

    \date{Received \today / Accepted \today}
    \pagerange{\pageref{firstpage}--\pageref{lastpage}} \pubyear{2013}

    \maketitle

    \label{firstpage}

%%%%%%%%%%%%%%%%%%%%%%%%%%%%%%%%%%%%%%%%%%%%%%%%%%%%%%%%%%%%%%%%%%%%%%%%%%%%%%%
    \begin{abstract}
	% context
	% aims
Asteroseismic investigations based on the wealth of data now available,
in particular from the CoRoT and {\it Kepler} missions, require a good
understanding of the relation between the observed quantities and the properties
of the underlying stellar structure.
\citet{Kallin2012} found a relation between their determination of the
asymptotic phase of radial oscillations in evolved stars and the 
evolutionary state, separating ascending-branch red giants from
helium-burning stars in the `red clump'.
Here we provide a detailed analysis of this relation, which is found to derive
from differences between these two classes of stars in the thermodynamic state
of the convective envelope.
There is potential for distinguishing red giants and clump stars
based on the phase determined from observations that are too short to allow
distinction based on determination of the period spacing for mixed modes.
The analysis of the phase may also point to a better understanding of
the potential for using the helium-ionization-induced acoustic glitch
to determine the helium abundance in the envelopes of these stars.
%\note [Promising rather a lot!]
    \end{abstract}


%%%%%%%%%%%%%%%%%%%%%%%%%%%%%%%%%%%%%%%%%%%%%%%%%%%%%%%%%%%%%%%%%%%%%%%%%%%%%%%
\section{Introduction}
%%%%%%%%%%%%%%%%%%%%%%%%%%%%%%%%%%%%%%%%%%%%%%%%%%%%%%%%%%%%%%%%%%%%%%%%%%%%%%%
\label{sect:intro}

%\note [A little about explosion of asteroseismology, with CoRoT and
%{\it Kepler}.]

The observations of stellar oscillations carried out by the 
CoRoT \citep{Baglin2009} and {\it Kepler} \citep{Boruck2010}
missions have revolutionized asteroseismology,
by providing extensive photometric data of high quality
for a large number and great variety of stars.
This includes a very substantial number of stars showing 
solar-like oscillations, i.e., modes that are intrinsically damped and
excited stochastically by vigorous near-surface convection.

A particularly interesting case are the red-giant stars, representing
late evolutionary phases of low- and moderate mass-stars
(see, for example, \citealp*{Kippen2012} for a general overview of
stellar evolution;
for a detailed discussion of the properties of red giants, see
\citealp*{Salari2002}).
After completing central hydrogen fusion stars evolve towards lower 
effective temperature $T_{\rm eff}$,
before ascending the red-giant branch (RGB) with increasing luminosity
and nearly constant $T_{\rm eff}$.
This evolution is driven by the formation of a helium core whose mass
grows through the continued hydrogen fusion in a shell around the core,
while the core radius decreases.
This is accompanied by a strong increase in the 
surface radius.
When the temperature in the core becomes sufficiently high,
fusion of helium to carbon and oxygen sets in.
For stars of mass lower than around $1.8 \Msun$
the pressure in the inert helium core
is dominated by degenerate electrons and helium fusion starts as a run-away
process, the so-called helium flash,
leading to a complex and poorly modelled evolution, before the
star settles down to quiescent helium fusion.
For higher-mass stars the core is non-degenerate, and helium fusion starts
in a more regular fashion.
The ignition of helium leads to an expansion of the core and
a corresponding contraction of the envelope, decreasing the surface
radius and hence the luminosity, with a slight increase in $T_{\rm eff}$.
In both cases the hydrogen-fusing shell outside the helium core continues
to contribute a substantial fraction of the total surface luminosity.
Stars in the core helium-fusion phase are often referred to as `clump' stars,
%\footnote{A distinction is often made between the red clump and the 
%so-called secondary clump,
%the latter consisting of higher-mass stars that did not ignite helium in 
%a flash. 
%We do not make this distinction here.}
from the location of such stars in the so-called `red clump'
in colour-magnitude diagrams of open clusters.
%\note [Do we need to distinguish between clump and secondary clump?
%SH: yes! JC-D and VSA: probably not.] 
For field stars, as opposed to cluster stars,
it is very difficult to distinguish 
between RGB and clump stars based just on their observed surface properties,
given the scatter induced, e.g., by differences in metallicity.
%\note [SH comment on effect of different metallicities; 
%JCD would have thought
%that it would be difficult to distinguish even if the star had the same
%metallicity, but different luminosity.]
%\note [I find that the discussion of metallicity is probably too
%complex to get into here.]

%\note [Introduce red giants as a particularly interesting case,
%distinguishing already here between ascending-branch stars and `clump' 
%stars.]

%\note [Importance of defining diagnostics that can characterize stellar
%properties, usefulness of asymptotics.
%At simplest (and most easily visible) level acoustically dominated modes,
%including radial modes.]

To ascertain the diagnostic potential of the observed oscillations
it is necessary to consider the properties of the observed oscillation modes
\citep*[for a detailed discussion of these properties and
asteroseismic diagnostics, see][]{Aerts2010}.
In red giants the compact core gives rise to a very high local gravitational
acceleration and hence buoyancy frequency in the deep interior of the star.
For non-radial modes, with spherical-harmonic degree $l$ greater than 0,
this leads to a dense spectrum of trapped internal gravity waves, or g modes,
in the core.
%\note [might need a general reference on oscillation properties]
In the envelope the modes behave as acoustic waves, or p modes.
Thus all non-radial modes have a mixed character, 
with g-mode behaviour in the core and p-mode behaviour in the envelope.
At certain frequencies there is a resonance with the acoustic behaviour
in the envelope, leading to modes of predominantly p-mode character,
with larger amplitudes in the envelope than in the core.
Owing to their lower inertia
such modes are easier to excite by the near-surface convection and hence
are generally more visible in the observations \citep[e.g.,][]{Dupret2009}.
To these non-radial modes must be added the spherically symmetric,
or radial, modes with $l = 0$ which are of purely acoustic character.

A major early breakthrough from the CoRoT observations of red giants
was the identification of non-radial acoustically dominated modes
\citep{DeRidd2009}.
Remarkably, this was soon followed by the detection of additional modes of
degree $l = 1$
with a substantial g-mode component \citep{Beck2011}.
Such modes are characterized by the period spacing between adjacent modes,
which is determined by the buoyancy frequency in the deep interior of the
star and hence provides a diagnostics of conditions in the core.
As a very important result it was shown by \citet{Beddin2011}
and \citet{Mosser2011a} that the period spacing is substantially
higher in clump stars than in RGB stars, providing a first clear 
observational separation, applicable to individual stars,
between these two classes.
This obviously requires a clear observational identification of the g-dominated
modes.
Such an identification has been possible on the basis of observations over
several years that were obtained during the nominal {\it Kepler} mission
\citep[e.g.,][]{Mosser2012, Silva2014}.
However, the shorter duration of the observations in the K2 extension of
{\it Kepler} \citep{Howell2014} and for the TESS mission \citep{Ricker2014}
makes unlikely a detailed analysis of g-dominated modes.

Here we concentrate on the acoustically dominated modes which, as mentioned
above, are probably those that are most easily observable.
Acoustic modes satisfy an approximate asymptotic relation which
can be written
\begin{equation}
\nu_{nl} \simeq \Delta \nu \left(n + {l \over 2} + \epsilon \right) 
- d_{0l} 
\label{eq:pasymp}
\end{equation}
\citep[e.g.,][]{Tassou1980},
where $\nu_{nl}$ is the cyclic frequency of the mode of radial order $n$
and degree $l$.
The separation between modes of the same degree and adjacent order
(the so-called large frequency separation) is approximately given by
\begin{equation}
\Delta \nu \simeq \left ( 2 \int_0^{R_*} {\dd r \over c} \right)^{-1} \; ,
\label{eq:asympsep}
\end{equation}
where $c$ is the adiabatic sound speed and the integral is over distance $r$
to the centre of the star, extending to a suitable value $R_*$ near the
surface of the star.
In equation (\ref{eq:pasymp})  $\epsilon$ is a phase which, 
as extensively discussed below, depends
on frequency and is probably predominantly determined by the 
properties of the near-surface layers of the star,
and $d_{0l}$ is a small correction.
This expression can relatively simply be derived for main-sequence stars
where $d_{0l}$ provides a useful diagnostic of the evolutionary stage
\citep[e.g.,][]{Christ1984, Christ1988, Ulrich1986}.
The applicability of equation (\ref{eq:pasymp}) 
to red giants is perhaps less obvious, in the light of their compact core,
but is clearly shown both by stellar models and by observations of red giants
\citep[e.g.,][]{DeRidd2009, Beddin2010, Huber2010},
%\note [a reference or two to pre-Mosser observations],
leading to the introduction of the so-called `universal pattern'
(\citealp{Mosser2011b}; see, however, \citealp{Stello2014}).
%Note that $\epsilon$ is generally thought to be predominantly determined
%by near-surface layers.

It follows from simple homology relations that $\Delta \nu$ scales as
the square root of the mean density of the star,
\begin{equation}
\Delta \nu \propto \left( {M \over R^3} \right)^{1/2} 
\label{eq:dnuscale}
\end{equation}
\citep{Ulrich1986}.
Thus determination of $\Delta \nu$ from observations provides a strong 
constraint on the global properties of the star,
although corrections to this scaling may be required,
as discussed, e.g., by \citet{White2011} and \citet{Miglio2012}.
However, as a result of the frequency dependence of $\epsilon$
the actual separation between adjacent modes,
\begin{equation}
\Delta \nu_{nl} \equiv \nu_{n+1\,l} - \nu_{nl}
\simeq \Delta \nu [1 + \epsilon(\nu_{n+1\,l}) - \epsilon(\nu_{nl})] \; ,
\label{eq:freqsep}
\end{equation}
in general differs from the asymptotic separation $\Delta \nu$.
This must be taken into account when the scaling relation (\ref{eq:dnuscale})
is applied to values of $\Delta \nu$ estimated from the observed frequencies.
On the other hand, the frequency dependence of $\Delta \nu_{nl}$ 
provides diagnostics of the outer layers of the star, including the
envelope helium abundance \citep[e.g.,][]{Miglio2010}. 
%\note [\citet{Broomh2014} use second differences and are discussed
%in the Discussion.]

%\note [Also a little on $\nu_{\rm max}$.]

A second important global asteroseismic diagnostics is the frequency
$\nu_{\rm max}$ where the power density is maximum.
This can be determined relatively accurately
%\note [there might be a point in mentioning that for red giants the
%power envelope is generally rather well defined]
by fitting, e.g., a Gaussian to the envelope of power.
It has been found observationally \citep[e.g.,][]{Brown1994, Stello2008},
with some theoretical support \citep{Belkac2011},
that $\nu_{\rm max}$ scales as the acoustic cut-off frequency in the stellar
atmosphere, leading to
\begin{equation}
\nu_{\rm max} \propto {M \over R^2} T_{\rm eff}^{-1/2} \; ,
\label{eq:numax}
\end{equation}
where $T_{\rm eff}$ is the effective temperature.
Given an observational determination of $T_{\rm eff}$ the radius and mass of
a star can be determined from observed values of $\Delta \nu$ and
$\nu_{\rm max}$ \citep[e.g.,][]{Kallin2010}.

%\note [The following should perhaps include a brief reference to White et al.]
%Alternative procedure based just on a few modes around $\nu_{\rm max}$,
%as introduced by Kallinger.
%Discuss results, including the perhaps surprising distinction between
%clump stars and RGB stars.]
%
%\note [Procedure for determining the phase for radial modes,
%following Kallinger.]

\citet{Kallin2012} emphasized the
ambiguity of the observational determination of 
$\Delta \nu$, depending on the selection of modes actually observed,
and argued that this might hide significant dependencies of the
inferred global asteroseismic parameters.
They noted that the large-scale variation of $\epsilon$,
reflected in the curvature of the ridges in an \'echelle diagram,
quite generally appears to have a local extremum near $\nu_{\rm max}$
and hence proposed to base the analysis on just a few radial modes
in the vicinity of $\nu_{\rm max}$.
This has the added advantage that these modes may be expected to be the most 
visible.
Specifically, they chose the radial mode, with order $n = \nc$,
closest to $\nu_{\rm max}$,
and defined the large separation by
\begin{equation}
\Delta \nu_{\rm c} = (\nu_{\nc + 1\,0} - \nu_{\nc - 1\,0})/2 \; .
\label{eq:delnuc}
\end{equation}
From this they also obtained, as a measure of the phase $\epsilon$,
\begin{equation}
\epsilonc = {\nu_{\nc 0} \over \Delta \nu_{\rm c}} \quad {\rm mod} \quad 1 \; .
\label{eq:epsilonc}
\end{equation}
%\note [Perhaps we can be more specific about the use of a weight function
%around $\nu_{\rm max}$ when discussing the model results in the next
%section; for the detailed analysis later we might get away with using this
%simple expression.]
Kallinger et al. used this procedure to analyse a large number of
red-giant stars for which the evolutionary phases (RGB or clump) were
already known from their mixed-mode period spacings.
They presented their results in a $(\Delta \nu_{\rm c}, \epsilonc)$ diagram.
Interestingly, they found evidence for a distinction in 
$\epsilonc$ between clump and RGB stars at fixed $\Delta \nu_{\rm c}$,
proposing the use of $\epsilonc$ as a diagnostics to identify clump stars
based on just the radial-mode frequencies.
Given the general preconception that $\epsilon$ is determined mainly by 
the surface layers, and the fact that the structural differences between
clump and RGB stars are predominantly in the core,
such a significant signature may be somewhat surprising.

%\note [Purpose of paper is to analyse this in a little more detail,
%understand better the diagnostic potential of the near-surface phase.]

Here we analyse the origin of the difference in behaviour
between the RGB and the clump model.
Section \ref{sec:model} presents results for stellar evolution models 
of various masses.
In Section \ref{sec:analysis} we analyse the relevant properties of stellar
oscillations for models on the RGB and in the clump
and relate the difference in $\epsilonc$ to the properties of the convection
zones of the two stars; a further analysis of the convection-zone properties
is provided in Appendix A.
Finally, Section \ref{sec:discuss} gives a discussion of the results and
summarizes our conclusions.

\section{Stellar modelling}
\label{sec:model}

%\note [Very briefly on the GARSTEC modelling.]

\begin{figure}
\centering
\includegraphics[width=\hsize]{\figdir/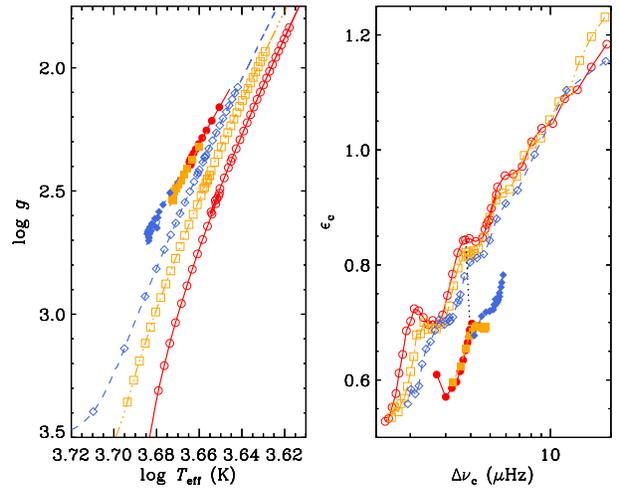}
\caption{
Left panel: evolution tracks on the ascending red-giant branch and the
clump for masses of
$1 \Msun$ (red circles), $1.5 \Msun$ (yellow squares) and 
$2 \Msun$ (blue diamonds),
computed with the {\sc GARSTEC} code.
Open symbols show models on the red-giant branch, and closed symbols show
helium-burning clump models.
Right panel: the corresponding large frequency separation $\Delta \nu_{\rm c}$
and phase $\epsilonc$, determined in the manner of \citet{Kallin2012}.
The models $\CMrgb$ and $\CMrc$
(cf. Table~\ref{tab:rcmodels}), which are analysed in detail below,
are indicated by being connected by the nearly vertical dashed line.
\label{fig:epsevol}
}
\end{figure}

We computed evolutionary tracks using the GARching STellar Evolution Code
\citep[{\sc GARSTEC};][]{Weiss2008}.
The input physics included the NACRE compilation of nuclear reaction rates
\citep{Angulo1999}, the \citet{Greves1998} solar mixture,
OPAL opacities \citep{Iglesi1996} for high temperatures supplemented by
low-temperature opacities from \citet{Fergus2005},
the 2005 version of the OPAL equation of state \citep*{Rogers1996}
and the mixing-length theory of convection as described by \citet{Kippen2012}.
Convective overshooting and diffusion of helium and heavy elements 
were not considered.
The parameters of the models were chosen to correspond to a calibrated 
solar model, including diffusion and settling of helium and heavy elements;
this resulted in the convective efficiency
$\alpha_{\rm MLT}=1.791$ and initial abundances by mass of hydrogen and
heavy elements of 0.712 and 0.0192.
The left panel in Fig.~\ref{fig:epsevol} shows resulting evolution tracks
for three different masses.
% with the ascending-branch red giants at slightly
%lower effective temperature and the helium-burning clump stars at slightly
%higher $T_{\rm eff}$.

Detailed calculations of individual frequencies were obtained for masses of
1.0, 1.5, and $2.0 \Msun$ using the Aarhus Adiabatic Oscillations Package
\citep[{\sc ADIPLS};][]{Christ2008}.
From the frequencies $\Delta \nu_{\rm c}$ and $\epsilonc$ were
determined as described above.
The results are shown in the right panel of Fig.~\ref{fig:epsevol}.
There is a clear distinction between the RGB models, at higher $\epsilonc$,
and the clump models, at lower $\epsilonc$,
as also inferred by \citet{Kallin2012}. 
%Additional models of $1.0 \Msun$ with different values of $\alpha_{\rm MLT}$
%were computed to test the effect in $\epsilonc$ of changes in the effective
%temperature, as described below.

%\note [Evolution tracks of the selected stars.]

%\mfigur{Epsilon-evolution.ps}{9.0cm}{fig:epsevol}
%    {\note [caption]
%}

%\note [Results.  Clear distinction between RGB and clump.]

\begin{figure}
\centering
\includegraphics[width=\hsize]{\figdir/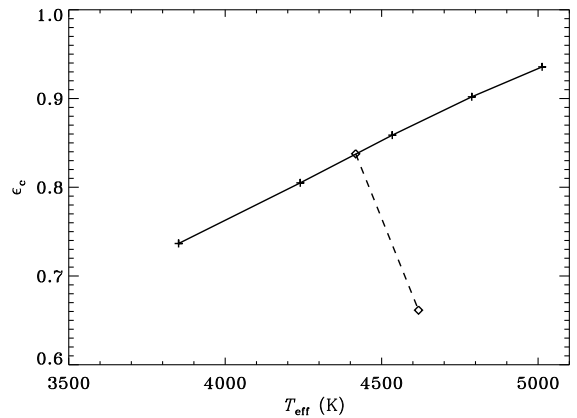}
\caption{The phase $\epsilonc$ as a function of effective temperature
for models at fixed radius ($10.69  \Rsun$).
Solid line: RGB models with $\alpha_{\rm ML} = 1$ (cooler) to 3 (hotter).
The diamonds connected by a dashed line show 
models at fixed $\alpha_{\rm ML}$ (cf. Table~\ref{tab:rcmodels}):
the RGB model $\CMrgb$ at high $\epsilonc$ and
the clump model $\CMrc$ at lower $\epsilonc$.
\label{fig:epsteff}
}
\end{figure}

In the following we discuss in detail the pair of $1 \Msun$ models,
$\CMrgb$ and $\CMrc$, at the same radius and approximately the same 
$\Delta \nu_{\rm c}$ on the RGB and in the clump, respectively.
The models are connected by the black dashed line in the
right panel of Fig. \ref{fig:epsevol}.
Some details about the models are provided in Table~\ref{tab:rcmodels}.

\begin{table}
\caption{Red-giant ($\CMrgb$) and clump ($\CMrc$) models, in a $1 \Msun$
evolution sequence, considered for detailed analysis of $\epsilonc$.
\label{tab:rcmodels}}
\begin{center}
\begin{tabular}{lcccc}
\hline\hline\noalign{\smallskip}
Model   &   $T_{\rm eff}$\,(K) & $R/\Rsun$ & $\Delta \nu_{\rm c}$ ($\mu$Hz) & 
$\epsilonc$ \\[5pt]
\hline\noalign{\smallskip}
$\CMrgb$ & 4417 & 10.69 & 3.760 & 0.8377 \\
$\CMrc$  & 4618 & 10.69 & 3.982 & 0.6615 \\
\noalign{\smallskip}
\hline\hline
\end{tabular}
\end{center}
\end{table}

\section{Analysis of the asymptotic phase}
\label{sec:analysis}

%\note [The following results are for the moment based on a pair models
%computed earlier (...andressa...).
%They should be replaced by a pair of models from Fig. \ref{fig:epsevol},
%which probably should be marked in that figure.]
%
Fig.~\ref{fig:epsevol} shows a significant difference between $\epsilonc$
between RGB and clump models at fixed $\Delta \nu$ for a given mass, and hence
at approximately fixed radius and surface gravity.
Here we aim at determining the dominant reason for this difference.
One possibility would be that $\epsilonc$ is directly related to the
effective temperature, which is the principal difference between the superficial
properties of the RGB and clump stars, 
on the assumption that $\epsilonc$ is predominantly determined by the stellar 
surface layers.
At the opposite extreme $\epsilonc$ might somehow be directly related to the
stellar core, the site of the most fundamental differences between the two
types of stars.

To test the assumption
that the difference in $\epsilonc$ is related to the difference 
in effective temperature between RGB and clump
we computed a set of evolution sequences for the
RGB, varying the mixing-length parameter $\alpha_{\rm ML}$
to change $T_{\rm eff}$ at given radius.
We selected models at fixed radius and evaluated $\epsilonc$ as discussed above.
In Fig. \ref{fig:epsteff} the results are compared with a pair of models
of the same mass and radius on the RGB and the clump,
discussed in more detail below.
It is evident that the change in $\epsilonc$ from RGB to clump is quite 
different from the effect expected simply from the change in $T_{\rm eff}$.
The effect on $\epsilonc$ must therefore have a more deep-seated origin.

The most dramatic difference between the structure of the RGB and clump
model is obviously in the core.
It was noted by \citet{Roxbur2000, Roxbur2003} that the phase in the
asymptotic expression for acoustic-mode frequencies does indeed contain
a contribution from the core of the star.
This can be analysed in terms of the eigenfrequency equation 
\begin{equation}
\omega_{nl} T_0 \simeq \pi \left(n + {l \over 2}\right)
+ \alpha_l(\omega_{nl}) - \delta_l(\omega_{nl}) \; ,
\label{eq:freqeq}
\end{equation}
where $\omega = 2 \pi \nu$ is the angular frequency, and 
\begin{equation}
T_0 = \int_0^{R_*} {\dd r \over c} 
\end{equation}
is the acoustic radius of the star;
note, from equation (\ref{eq:asympsep}), that $\Delta \nu = 1/(2 T_0)$.
The core phase $\delta_l(\omega)$ and the envelope phase
$\alpha_l = \alpha(\omega)$
can be obtained as functions of 
frequency by fitting partial solutions to the oscillation equations
(i.e., solutions that do not satisfy all the boundary conditions)
to the relevant asymptotic expressions. 
In the case of $\alpha$ this was developed in some detail 
by \citet{ChristPH1992}
\citep[see also][for a slightly different treatment of the envelope phase]
{Brodsk1988, Voront1989, Roxbur1996}.
As indicated, the envelope phase is independent of $l$ for low-degree modes.

As also noted by \citet{Kallin2012}
equation (\ref{eq:freqeq}) is clearly equivalent to equation (\ref{eq:pasymp}), 
with
\begin{equation}
\epsilon = {1 \over \pi} (\alpha - \delta_0) \; ,
\label{eq:asepsilon}
\end{equation}
and
\begin{equation}
d_{0l} = {\Delta \nu \over T_0} (\delta_l - \delta_0) \; .
\label{eq:d0l}
\end{equation}
Equation (\ref{eq:asepsilon}) shows explicitly the contribution,
through $\delta_0$, to $\epsilon$ from the inner parts of the star.
To relate $\epsilonc$ as defined by \citet{Kallin2012} to the properties
of $\epsilon$ we note from equation (\ref{eq:pasymp}),
with $l = 0$ and $d_{0l} = 0$,
equation (\ref{eq:delnuc}) and equation (\ref{eq:epsilonc}), that
\begin{equation}
\epsilonc = {\nc + \epsilon(\nu_{\nc 0}) \over  
 1 + [\epsilon(\nu_{\nc+1\,0}) - \epsilon(\nu_{\nc-1\,0})]/2 }
\quad {\rm mod} \quad 1 \; .
\label{eq:epsepsc}
\end{equation}
Thus $\epsilonc$ depends on both $\epsilon(\nu)$ and its variation with 
frequency.

\begin{figure}
\centering
\includegraphics[width=\hsize]{\figdir/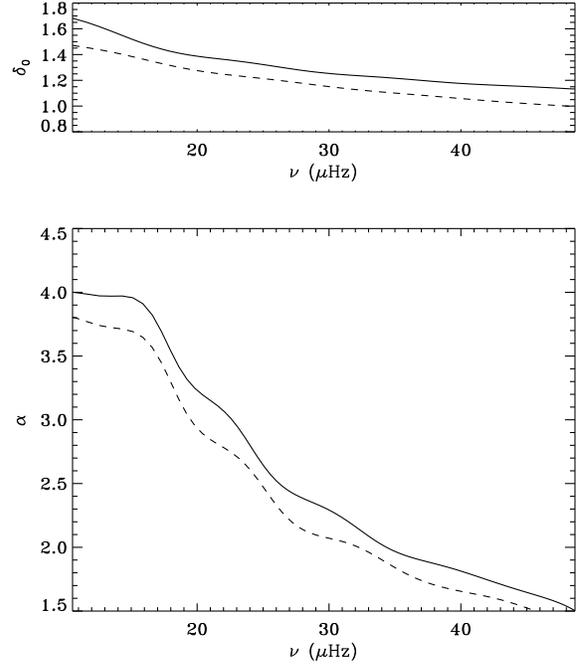}
\caption{Core phase $\delta_0$ for radial waves (upper panel)
and envelope phase $\alpha$ (lower panel), as functions of frequency,
for two models at the same radius.
Solid lines show results for the red-giant model $\CMrgb$ and
dashed lines results for the clump model $\CMrc$
(cf. Table~\ref{tab:rcmodels}).
\label{fig:clumpphs}
}
\end{figure}

\begin{figure}
\centering
\includegraphics[width=\hsize]{\figdir/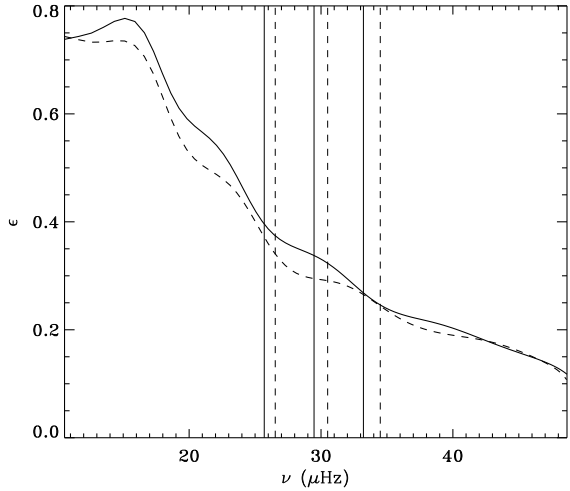}
\caption{Combined phase $\epsilon$ as a function of frequency,
for two models at the same radius.
The solid line shows results for the red-giant model $\CMrgb$ and
the dashed line results for the clump model $\CMrc$
(cf. Table~\ref{tab:rcmodels}).
The solid and dashed vertical lines
mark the frequencies used for determination of $\epsilonc$ in
models $\CMrgb$ and $\CMrc$, respectively.
\label{fig:clumpeps}
}
\end{figure}

The phase functions were determined by matching computed partial solutions
to the relevant asymptotic expressions, in terms of the function
\begin{equation}
\psi(r) = \frac{(\rho c)^{1/2}}{r} p' \; ,
\label{eq:asfunction}
\end{equation}
where $\rho$ is density and $p'$ is the Eulerian pressure perturbation.
In the inner part of the model the relevant form is
\begin{equation}
\psi_{\rm c}(r)
\simeq A_{\rm c} \sin\left(\omega \tilde\tau - \pi l/2 
+ \delta_l(\omega)\right) \; ,
\label{eq:innerfit}
\end{equation}
where $\psi_{\rm c}$ is obtained from a partial solution satisfying just the
central boundary conditions;
here
\begin{equation}
\tilde \tau = \int_0^r {\dd r' \over c} \qquad
\label{eq:acradius}
\end{equation}
is the acoustic distance to the centre.
The term in $l$ is included to acknowledge the behaviour of the solution
in main-sequence stars, owing to the singular point at $r = 0$
\citep[see][]{Roxbur2003}.
In the outer parts of the model the relevant form is
\begin{equation}
\psi_{\rm e}(r)
\simeq A_{\rm e} \sin\left(\omega \tau - \alpha(\omega)\right) \; ,
\label{eq:outerfit}
\end{equation}
where $\psi_{\rm e}$ is obtained from a partial solution satisfying just the
surface boundary conditions and
\begin{equation}
\tau = \int_r^{R_*} {\dd r' \over c} 
\label{eq:acdepth}
\end{equation}
is acoustic depth.
Note that $\delta_l(\omega)$ and $\alpha(\omega)$ are defined for {\it any}
frequency $\omega$.
The eigenfrequency equation, equation (\ref{eq:freqeq}), follows from
equations (\ref{eq:innerfit}) and (\ref{eq:outerfit}) by demanding that
$\psi_{\rm c}$ and $\psi_{\rm e}$ represent the same solution which is 
continuous and with continuous first derivative at some suitable fitting
point $r_{\rm f}$.

\begin{figure}
\centering
\includegraphics[width=\hsize]{\figdir/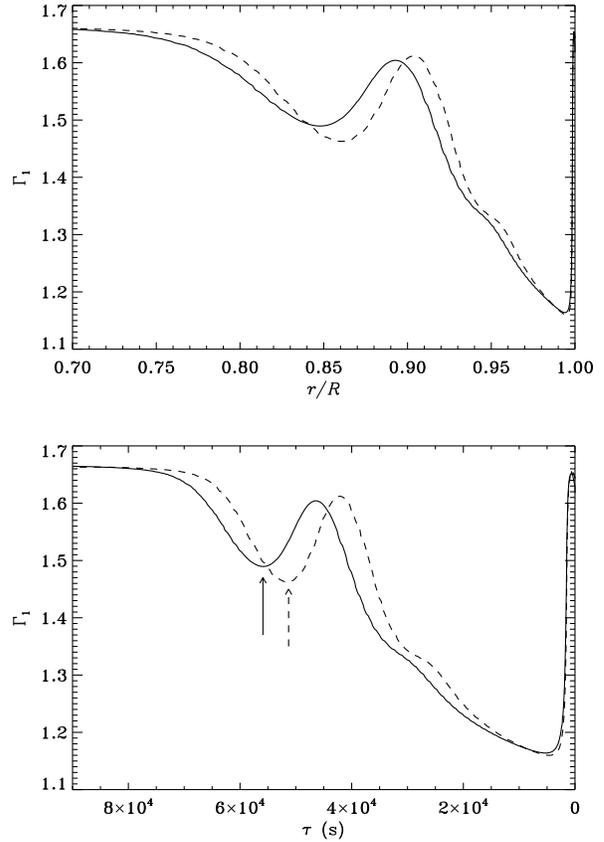}
\caption{Adiabatic compressibility $\Gamma_1$ 
for two models at the same radius.
The solid line shows the red-giant model $\CMrgb$ and
the dashed line the clump model $\CMrc$ (cf. Table~\ref{tab:rcmodels}).
In the upper panel $\Gamma_1$ is shown against fractional radius,
while in the lower panel the abscissa is acoustic depth
(cf. equation \ref{eq:acdepth}).
The solid and dashed arrows mark the dips in $\Gamma_1$ caused by 
the second helium ionization in models $\CMrgb$ and $\CMrc$, respectively.
\label{fig:gamma1}
}
\end{figure}

The properties of $\alpha$ and $\delta_0$ and their effect on 
$\epsilonc$ can be illustrated by considering 
models $\CMrgb$ and $\CMrc$ (cf. Table~\ref{tab:rcmodels}),
utilizing the fact that from equation (\ref{eq:asepsilon}) $\epsilon$
can be obtained as a continuous function of frequency.
The core and envelope phases for these models are illustrated in
Fig. \ref{fig:clumpphs} and the resulting $\epsilon$ is shown in
Fig. \ref{fig:clumpeps}; 
here the phases were determined by fitting the solutions to the
relevant asymptotic expressions
(equations \ref{eq:innerfit} and \ref{eq:outerfit}) near $r/R = 0.75$,
chosen to be below the region of varying adiabatic compressibility $\Gamma_1$
(cf. Fig.  \ref{fig:gamma1}).
Interestingly, the overall difference between the two models is
similar for $\delta_0$ and $\alpha$, and hence the scale of the difference in
$\epsilon$ is modest.
However, there are clear differences in the frequency dependence of
$\alpha$ and hence $\epsilon$.
This has a substantial effect on $\epsilonc$ through the denominator in
equation (\ref{eq:epsepsc}).

To make this explicit and separate the contributions from $\delta_0$ 
and $\alpha$ we first linearize the change in $\epsilonc$ in the changes in
$\epsilon$.
We introduce $\epsilon_0 = \epsilon(\nu_{\nc\,0})$,
$\epsilon_+ = \epsilon(\nu_{\nc+1\,0})$ and
$\epsilon_- = \epsilon(\nu_{\nc-1\,0})$.
Then according to equation (\ref{eq:epsepsc})
\begin{equation}
\epsilonc = {\nc + \epsilon_0 \over  
 1 + (\epsilon_+ - \epsilon_-)/2 } \quad {\rm mod} \quad 1 = 
{\nc + \epsilon_0 \over  1 + (\epsilon_+ - \epsilon_-)/2 } - \nc \; ,
\label{eq:epsepsc1}
\end{equation}
where the second equality was found to be satisfied for the models 
considered here, or
\begin{equation}
\epsilonc = {\epsilon_0 - \nc ( \epsilon_+ - \epsilon_-)/2 \over  
 1 + (\epsilon_+ - \epsilon_-)/2 } \; .
\label{eq:epsepsc2}
\end{equation}
Linearizing this in small changes $\delta \epsilon$ to $\epsilon$ we obtain
\begin{equation}
\delta \epsilonc \simeq 
{\delta \epsilon_0 \over  1 + (\epsilon_+ - \epsilon_-)/2 } - 
{(\epsilon_0 + \nc) ( \delta \epsilon_+ - \delta \epsilon_-)/2 \over  
[ 1 + (\epsilon_+ - \epsilon_-)/2]^2 } \; .
\label{eq:depsilonc}
\end{equation}
It is obvious that 
$\delta \epsilon = \pi^{-1} (\delta \alpha - \delta (\delta_0))$. 
Thus equation (\ref{eq:depsilonc})  can be used to estimate the separate 
contributions from the core, through $\delta (\delta_0)$,
and from the envelope, through $\delta \alpha$.
%\note [The notation $\delta \delta_0$ is not great, actually!]

Some numerical results are shown in Table~\ref{tab:epsdif},
comparing a pair of RGB models with different $T_{\rm eff}$ 
(cf. Fig. \ref{fig:epsteff}) with the RGB and clump pair.
The linearized expression in equation (\ref{eq:depsilonc}) recovers the
actual difference in $\epsilonc$ to within 2 per cent.
Also, the envelope contribution is completely dominant,
by more than a factor of 6 in the RGB - clump case,
thus confirming what one might have naively expected.
This dominance is even stronger, obviously, for models along the sequence
of RGB models with varying $\alpha_{\rm ML}$, given that in this case there
is little variation in the internal structure.

\begin{table}
\caption{Comparison of phases in two pairs of $1 \Msun$ models,
all with radius $10.69 \Rsun$.
The central column compares two RGB models of different mixing length
(cf. Fig.  \ref{fig:epsteff}), whereas the right-hand column compares
the RGB model $\CMrgb$ (1) and the clump model $\CMrc$ (2).
The phases $\epsilonc$ were determined from equation (\ref{eq:epsepsc}) based on
the phase function $\epsilon(\nu)$,
and linearized difference was obtained from equation (\ref{eq:depsilonc}),
with corresponding determination of the $\delta$ (core) and $\alpha$ 
(envelope) contributions.
\label{tab:epsdif}}
\begin{center}
\begin{tabular}{lcc}
\hline\hline\noalign{\smallskip}
   &   RGB,  &  RGB, Clump \\
   & varying $\alpha_{\rm ML}$ & \\[5pt]
\hline\noalign{\smallskip}
$T_{\rm eff}(1)$\,(K) & 4417 & 4417 \\
$T_{\rm eff}(2)$\,(K) & 4788 & 4618 \\
$\epsilonc(1)$ & 0.8376  & 0.8376 \\
$\epsilonc(2)$ & 0.8994  & 0.6615  \\
Difference     & 0.0618  & $-0.1772$ \\
Linearized     &       &          \\
difference     & 0.0614  & $-0.1801$  \\
$\delta$ contribution & $-0.0008$ & 0.0278 \\
$\alpha$ contribution & 0.0622 & $-0.2079$ \\
\noalign{\smallskip}
\hline\hline
\end{tabular}
\end{center}
\end{table}

%\note [A little about what dominates the difference;
%somehow argue that the important aspect is the wiggles in $\epsilon(\nu)$.
%Also note the amplification coming from $\nc$ (8 in the present case).]
%
To understand the origin of the strong difference in
behaviour as a function of $T_{\rm eff}$ seen in Fig. \ref{fig:epsteff}
we note that the oscillations in $\epsilon$ as a function of frequency
(cf. Fig. \ref{fig:clumpeps}) are strongly reminiscent of the effect on
the frequencies
of the acoustic glitch associated with the second helium ionization zone,
induced by the variation in $\Gamma_1$
(\citealp*{Gough1990, Voront1991}; for a detailed analysis of the effects
of the helium glitch, see \citealp{Verma2014}).
%\note [And in fact it might be good to relate Vorontsov functions 
%to $\epsilon$ somewhere -- perhaps even in an appendix.]
A glitch located at a depth $\tau_{\rm g}$ gives rise to a variation in
the frequency, and correspondingly in $\epsilon$,
of the form $\sin(2 \omega \tau_{\rm g})$ and hence with a
`period' in cyclic frequency of $1/(2 \tau_{\rm g})$.
Fig.~\ref{fig:gamma1} shows $\Gamma_1$ in the two models,
clearly showing that the acoustic glitch from the second helium ionization zone
in the clump model is somewhat shallower in acoustic depth and
of larger amplitude than in the RGB model.
Correspondingly, at least in a qualitative sense, the oscillatory variation of
$\epsilon$ in the clump model (cf. Fig. \ref{fig:clumpeps}) is stronger and
with a longer period than in the RGB model, causing differences in
the behaviour around $\nu_{\rm max}$ and hence in $\epsilonc$.
%\note [Quantitatively it does not seem to work so well; this may need further
%study, now or later.]
%The acoustic depth of the local minimum in $\Gamma_1$ 
%\note [could mark in the figure, quote values]
%would be expected to yield an oscillatory variation in the frequency
%and hence in $\epsilon$ with a `period' of \note [quote values],
%closely matching the behaviour seem in Fig. \ref{fig:clumpeps}
%\note [we hope].
%\note [In fact we may need to say a little more about the oscillations in
%$\epsilon$, corresponding to the glitch, and the relation to the acoustic
%depth.
%This may also depend somewhat of the success of explaining the difference
%in $\epsilonc$ in terms of the difference in acoustic depth.
%Again, probably something to defer to a later study.]

\begin{figure}
\centering
\includegraphics[width=\hsize]{\figdir/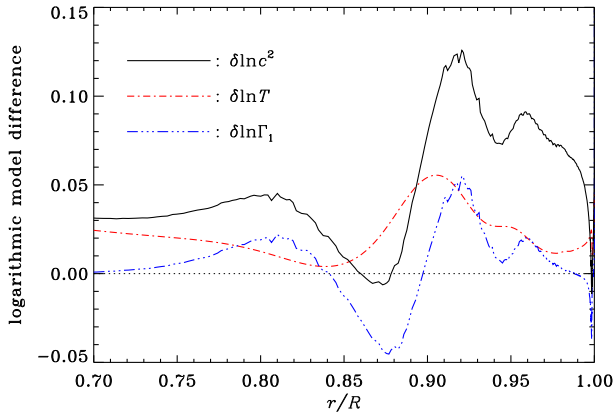}
\caption{Differences in natural logarithm between quantities in
the clump model $\CMrc$ and the RGB model $\CMrgb$ at fixed fractional radius,
in the sense $\CMrc - \CMrgb$.
Solid line: $c^2$; dot-dashed line: $T$; triple-dot-dashed line: $\Gamma_1$.
\label{fig:moddif}
}
\end{figure}

%\note [Go on to discuss underlying reason for this difference in terms of
%different adiabat (which requires a little understanding).]

As a further illustration of the origin of the different properties
of $\epsilonc$ between the clump and RGB model,
Fig.~\ref{fig:moddif} shows differences between these two models.
The direct effect on $\epsilon(\nu)$ is undoubtedly related to the difference
in $c^2$ which, assuming the ideal gas law and given that the composition
is essentially the same, is proportional to $\Gamma_1 T$.
Thus the sound-speed difference arises from the sum of the differences
in $T$ and $\Gamma_1$, with the sharp feature dominated by 
$\Gamma_1$, as assumed in the discussion of the glitches above.
These differences arise from a difference in the thermodynamic state
between the convective envelopes in the two models, related to a substantial
difference in the adiabatic constant, characterized by
$p/\rho^{\Gamma_1}$ in the deeper parts of the convection zone.
In particular, in the bulk of the convection zone the density
at fixed fractional radius is lower by about a factor 0.4 in 
model $\CMrc$, compared with model $\CMrgb$.
This leads to an increase in the
degree of ionization in the clump model relative to the RGB model,
and hence to the outward shift in the location of helium ionization,
reflected in $\Gamma_1$ (cf. Fig. \ref{fig:gamma1}).
The difference in density appears to be related to
the fact that a larger fraction of the mass is contained in the compact helium
core in the clump model;
we discuss this in more detail in the Appendix.
In the corresponding pair of RGB models with different mixing length
the differences in the adiabatic constant, and other thermodynamic quantities,
are minimal.
The increase in the effective temperature at fixed radius, caused
by the increase in the mixing length, is accompanied by a {\it decrease} in the
temperature in the bulk of the convection zone; 
this is a consequence of the increase in the convective efficacy and hence
a decrease in the superadiabatic temperature gradient.
This leads to a shift of the second helium ionization zone to slightly 
greater acoustic depth, causing the modest increase in $\epsilonc$
(cf. Fig.~\ref{fig:epsteff} and Table~\ref{tab:epsdif}).

The analysis in the Appendix provides a more detailed understanding of the
properties of the convective envelope and its relation to the core of the 
star, expressed in terms of the adiabatic constant $K$ in the bulk of the 
convection zone.
An interesting result is the dependence on $K$ of the temperature and 
sound speed in the deeper parts of the convection zone.
This appears to be related, at least in a qualitative sense, to a feature
noted by \citet{Miglio2012} in fitting global asteroseismic
observations for RGB and clump stars;
Miglio et al. showed that differences in the sound-speed structure between these
two classes of stars have a significant effect when using the scaling
relation (\ref{eq:dnuscale}) in such fits.

\section{Discussion and conclusions}
\label{sec:discuss}

The properties of the phase $\epsilonc$, determined according to 
\citet{Kallin2012} from radial modes near the frequency of 
maximum oscillation power,
provide an interesting diagnostic to separate stars on the red-giant branch
from clump stars.
We have demonstrated that this effect is not a direct result of the
substantial differences in core structure between these two classes of stars.
Instead, it is caused by differences in the thermodynamic state of the
convection zone, shifting the location in acoustic depth of the
acoustic glitch caused by the second helium ionization zone.
We have analysed this in terms of the phase function $\epsilon(\nu)$ which,
as indicated, is defined for any frequency.
The shift in the acoustic glitch causes a
change in the oscillatory behaviour of $\epsilon$ and hence,
as a result of the local nature of the definition used by
Kallinger et al., a change in $\epsilonc$.
The effect on $\epsilon$ of these changes in the model
structure might instructively be analysed in terms of the 
kernels for the phase function introduced by \citet{ChristPH1992}.
%\note [SH and YE: please check the following bit.]
We note that Vrard et al. (in preparation) fitted the variation
in the separation  of the radial modes as a function of frequency 
to take account of the glitch due to the signature of the discontinuity
from the second helium ionization zone. 
They used the phase of this oscillatory signal as a function of
$\nu_{\rm max}$ to explain the evolutionary classification based on
$\epsilonc$ proposed by \citet{Kallin2012}.
The observational approach by Vrard et al. is closely related to the
theoretical analysis performed here, reaching consistent conclusions.

%\note [The following pararaph from Saskia (8/8/14) has been edited a little.]

It is of obvious potential interest to use $\epsilonc$ as a diagnostic of
evolutionary state for shorter data sets where the mixed modes, and hence
the g-mode period spacing, cannot be obtained.
We have estimated the expected uncertainty in $\epsilonc$ in 50-day 
observations, using standard error propagation,
based on the scatter in the results of $\Delta \nu_{\rm c}$ 
obtained using the method described by \citet{Kallin2012}
for twelve 50-day datasets per star for nearly 1000 stars 
\citep{Hekker2012}.
This expected uncertainty is typically of order 0.05 -- 0.07 for stars
in the frequency range of the red clump.
We expect this to increase for even shorter datasets.
As shown by \citet{Kallin2012} the width of the scatter in $\epsilon_{\rm c}$
for RGB stars is of the order of 0.1 while for clump stars the spread in
$\epsilon_{\rm c}$ is 0.2 to 0.3 (their Fig. 4).
The value of the expected uncertainty is such that we expect a significantly
larger confusion rate at the boundary of the RGB and RC as indicated by the 
(arbitrary) limit defined by \citet{Kallin2012} (dotted line in their Fig. 4).
This confusion has not been defined by Kallinger et al.,
but is expected to be (at least partly) due to larger intrinsic uncertainties
caused by the stochastic excitation of the oscillations which affects
results from short datasets.
%\note [I added the following concluding sentences; please check!]
The estimated uncertainty is not negligible compared with the difference
at fixed $\Delta \nu_{\rm c}$, of order 0.2, in $\epsilonc$ between the RGB and
the clump stars (cf. Fig. \ref{fig:epsevol} and Table~\ref{tab:rcmodels});
even so, we expect that determination of $\epsilonc$ from at least 50 days 
of observation will provide some separation between RGB and clump stars.
For even shorter datasets, such as obtained over most of the sky with TESS,
such a separation may be questionable but should certainly be investigated.

The variation with frequency of $\epsilon$ is directly related to the 
individual frequency separation $\Delta \nu_{nl}$ between modes
of adjacent orders (cf. equation \ref{eq:freqsep}).
\citet{Miglio2010} detected this variation in a red giant observed by
CoRoT and used the period of the variation, reflecting the acoustic depth
of the variation in $\Gamma_1$ caused by the second helium ionization zone,
as a diagnostics of the stellar properties.
Miglio et al. also noted that the amplitude of the variation
in principle provides a measure of the envelope helium abundance,
although the available data did not allow a meaningful determination.
The oscillatory behaviour of the envelope phase, or various combinations
based on it, have been used extensively for estimating the solar envelope
helium abundance, starting with \citet{Voront1991}.
%\note [Probably a few more selected references here.]
\citet{PerezH1994a} pointed out that a cleaner measure could be obtained by
filtering the phase function to take out the slowly varying part;
this was used by \citet{PerezH1994b} to estimate the solar helium abundance,
while \citet{PerezH1998} evaluated the diagnostic potential in 
main-sequence stars showing solar-like oscillations.
\citet{Houdek2007} investigated the diagnostics
of the solar envelope on the basis of second differences of frequencies
at fixed degree, to isolate the oscillatory component of the variation,
and applying an asymptotic analysis of the near-surface behaviour of the
oscillations to characterize the effect of the helium abundance on the
seismic signature.

A detailed analysis of the potential for diagnostics of the helium ionization
zone based on acoustic-mode frequencies in red giants was carried out by
\citet{Broomh2014},
on the basis of fits to second differences of frequencies similar to those
of \citet{Houdek2007}. 
Broomhall et al.\ noted the difficulties arising from 
the sparcity of the acoustically-dominated modes, and the complexity induced
for dipolar modes by the mixed behaviour.
Their results showed, as also found by \citet{Miglio2010}, that the 
acoustic depth of the helium ionization could be determined reasonably 
reliably,
whereas the amplitude of the signal, reflecting the abundance of helium,
was difficult to obtain with an accuracy providing a significant abundance
determination, even with the nearly four years of data provided by the
{\it Kepler} mission.
It is possible that a more detailed analysis of the envelope phase,
perhaps involving the filtering proposed by \citet{PerezH1994a},
could lead to a more robust determination of the abundance.
This probably deserves further investigation.

\section*{Acknowledgements}
We thank G. Houdek for useful discussions, and A. Weiss and the
anonymous referee for comments on earlier versions of the manuscript
which helped improving the presentation.
We gratefully acknowledge the financial and organizational support
from the Lorentz Center, Leiden, through the workshop `Asteroseismology
in red-giant stars', where this project was initiated.
Funding for the Stellar Astrophysics Centre is provided by 
the Danish National Research Foundation (Grant DNRF106).
The research is supported by the European Research Council through
the ASTERISK project
(ASTERoseismic Investigations with SONG and Kepler;
Grant agreement no 267864) and the StellarAges project
(Grant agreement no 338251).
YE acknowledges support from the UK Science and Technology Facilities Council
(STFC).
This research has made use of NASA's Astrophysics Data System.

%\bibliography{bibs/conv,bibs/opac,bibs/eos,bibs/starmod,bibs/seism,bibs/obs,bibs/math,bibs/staratm}
%\input{alfa-fit.bbl}

%\bsp

\appendix

\section{Properties of convective envelopes}

\begin{figure}
\centering
\includegraphics[width=\hsize]{\figdir/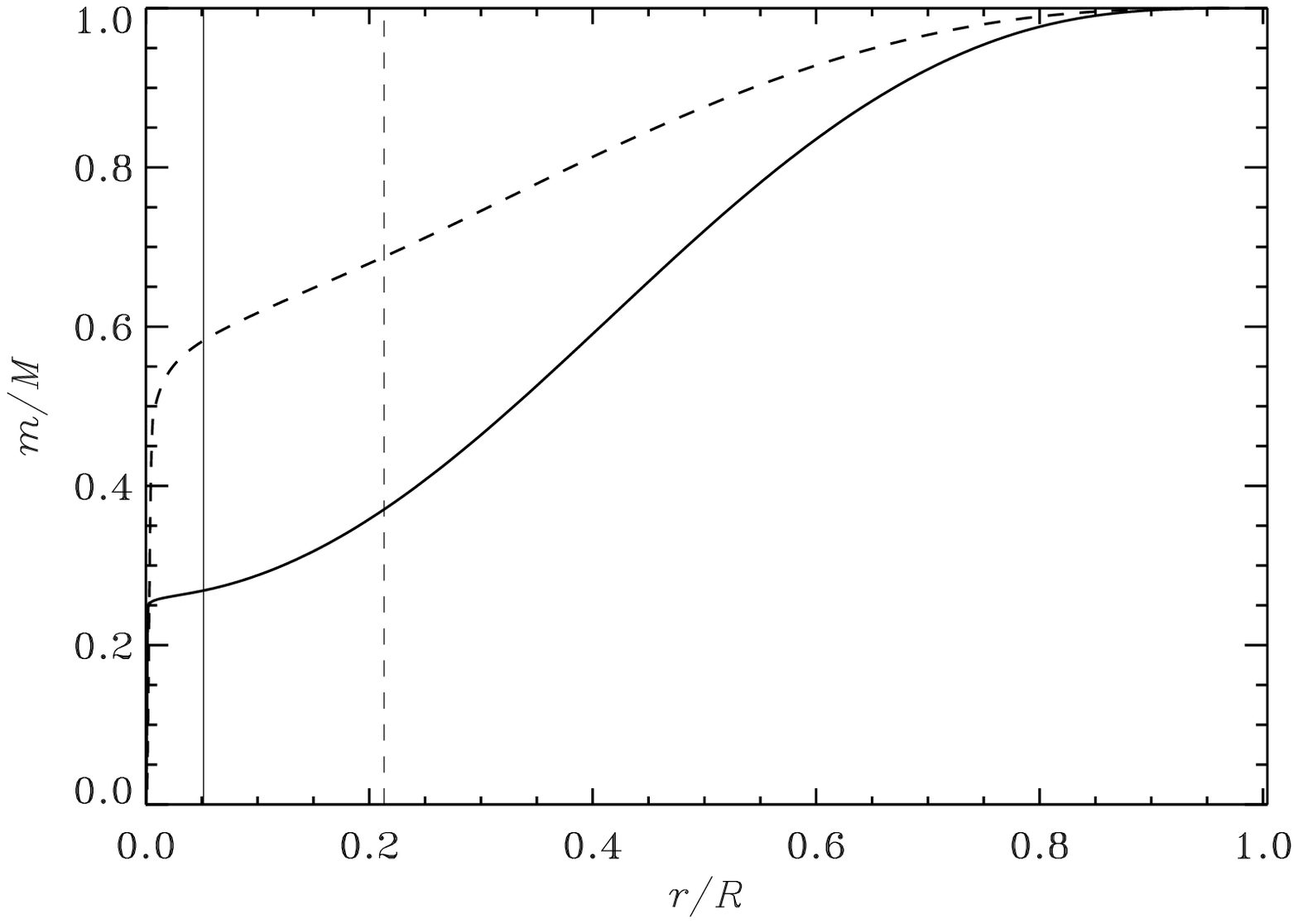}
\caption{Fractional mass as a function of fractional radius in the
RGB model $\CMrgb$ (solid line) and the clump model $\CMrc$
The vertical solid and dashed lines mark the base of the convective envelope
in the RGB and the clump models, respectively.
\label{fig:massfrac}
}
\end{figure}

\begin{figure}
\centering
\includegraphics[width=\hsize]{\figdir/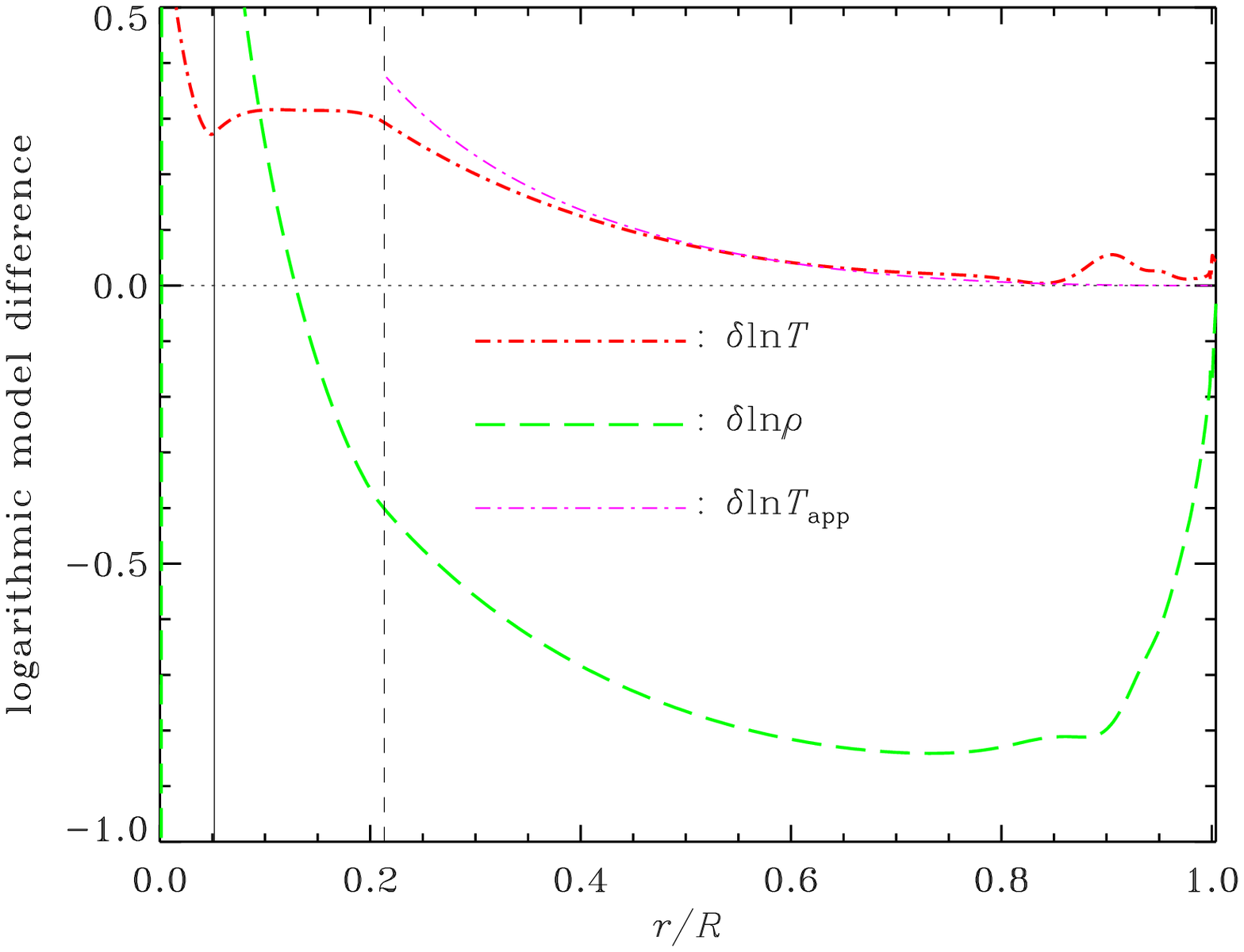}
\caption{Differences in natural logarithm between quantities in
the clump model $\CMrc$ and the RGB model $\CMrgb$ at fixed fractional radius,
in the sense $\CMrc - \CMrgb$.
Dashed line: $\rho$; dot-dashed line: $T$.
The vertical solid and dashed lines mark the base of the convective envelope
in the RGB and the clump models, respectively.
The thin dot-dashed line, confined to the convective envelope
in the clump model, shows the temperature difference
$\delta \ln T_{\rm app}$
estimated from the approximation corresponding to equation (\ref{eq:usoli}).
\label{fig:moddifrho}
}
\end{figure}

It is of obvious interest to obtain a better understanding of the differences
between the RGB and clump model in the structure of the convective envelope,
reflected in Fig. \ref{fig:moddif}.
As far as the envelope is concerned, an important difference between the models
is that in the clump star a larger fraction of the mass is contained in the 
helium core and the deep interior, as illustrated in Fig. \ref{fig:massfrac}.
Correspondingly, less mass is contained in the convective envelope at a
given radius in the clump model, obviously corresponding to a lower density.
This is made explicit in Fig. \ref{fig:moddifrho}, showing that the 
density at fixed radius in model $\CMrc$ is lower than in model $\CMrgb$ 
by more than a factor 2 in much of the convective envelope;
in contrast, the temperature difference is modest.
It is probably this difference in density that mainly affects the ionization
state of helium, leading to the differences in $\Gamma_1$ and sound speed
illustrated in Fig. \ref{fig:moddif}.

In the following we analyse these differences in more detail 
\citep*[see also, for example,][]{Christ1992, Christ1997}.
We note that the bulk of the 
convective envelope is adiabatically stratified.
Neglecting furthermore the variation in $\Gamma_1$ in the ionization zones
and taking $\Gamma_1 \simeq \gamma$ to be constant,
$p$ and $\rho$ are related by
\begin{equation}
p = K \rho^\gamma \; ,
\label{eq:adiab}
\end{equation}
defining the adiabatic constant $K$.
Introducing $u = p / \rho$, it follows from the equation of hydrostatic
equilibrium,
\begin{equation}
{\dd p \over \dd r} = - {G m \rho \over r^2} \; ,
\label{eq:hydeq}
\end{equation}
where $G$ is the gravitational constant,
that 
\begin{equation}
{\dd u \over \dd r} \simeq - {\gamma - 1 \over \gamma} {G m \over r^2} \; .
\label{eq:hydequ}
\end{equation}
In the outer parts of the convective envelope we can neglect the variation in
mass and take $m \simeq M$.
Then equation (\ref{eq:hydequ}) can be immediately integrated, to yield
\begin{equation}
u \simeq {\gamma -1 \over \gamma} {G M \over \Rs} 
\left( {\Rs \over r} - 1\right)
= {\gamma -1 \over \gamma} {G M \over \Rs} \xi \; ,
\label{eq:usol}
\end{equation}
where $\Rs \simeq R$ is a suitable reference radius and $\xi = \Rs / r - 1$.
It follows that the squared sound speed is 
\begin{equation}
c^2 = \gamma u \simeq (\gamma - 1) {G M \over \Rs} \xi \; .
\label{eq:csol}
\end{equation}
Also, assuming the ideal gas law, the temperature is approximated by
\begin{equation}
T \simeq {\mu m_{\rm u} \over k_{\rm B}} {p \over \rho}
\simeq {\gamma -1 \over \gamma}{\mu m_{\rm u} \over k_{\rm B}} 
{G M \over \Rs} \xi \; ,
\label{eq:tsol}
\end{equation}
where $\mu$ is the mean molecular weight, $m_{\rm u}$ is the atomic mass 
unit and $k_{\rm B}$ is Boltzmann's constant.
Finally, it follows from equations (\ref{eq:adiab}) and (\ref{eq:usol}) that
\begin{equation}
\rho \simeq \rho_0 \xi^n \; ,
\label{eq:rhosol}
\end{equation}
where $n = 1/(\gamma - 1)$ is the polytropic index and
\begin{equation}
\rho_0 = K^{-n} 
\left({\gamma -1 \over \gamma} {G M \over \Rs} \right)^n \; .
\label{eq:rhozero}
\end{equation}
For the fully ionized ideal gas relevant here $\gamma = 5/3$,
and hence $n = 3/2$.

In the comparison of models $\CMrgb$ and $\CMrc$, which have approximately
the same mass and radius,
equation (\ref{eq:tsol}) predicts that there should be little difference
in temperature.
As shown in Fig. \ref{fig:moddifrho} this is indeed approximately the case
in the outer parts of the convective envelope,
although with modest differences related to variations induced by the
helium ionization zone.
On the other hand, equation (\ref{eq:rhosol}) predicts a difference 
$\delta \ln \rho \simeq -n \delta \ln K$, given by the difference in 
the adiabatic constant, which is independent of position, which again is 
approximately satisfied in the outer parts of the convection zone,
as shown in Fig. \ref{fig:moddifrho}.

\begin{figure}
\centering
\includegraphics[width=\hsize]{\figdir/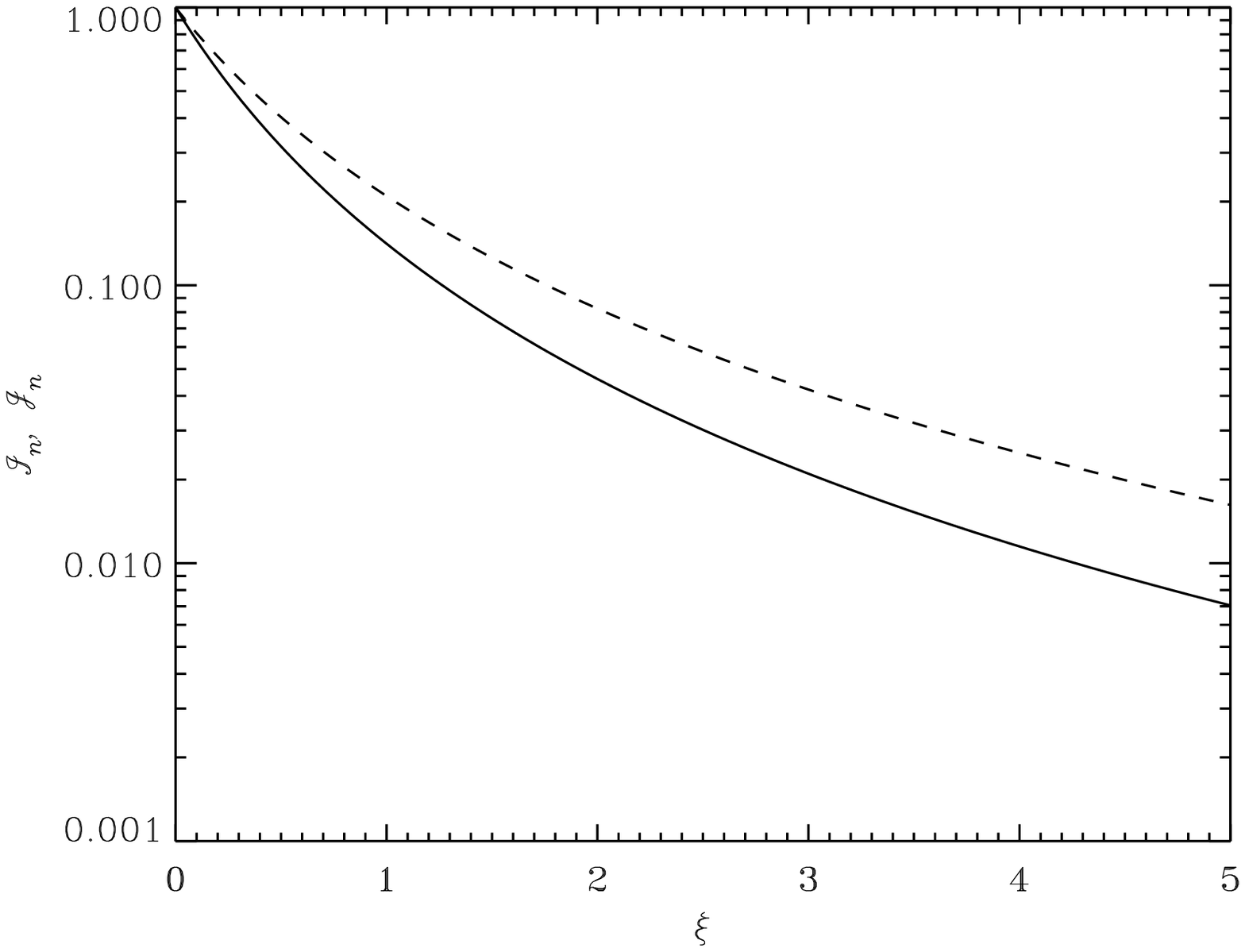}
\caption{The functions $\CI_n(\xi)$ (solid line) and $\CJ_n(\xi)$ (dashed line),
defined by equations (\ref{eq:curli}) and (\ref{eq:curlj}), for $n = 3/2$,
corresponding to the value $\gamma = 5/3$ relevant for a fully ionized
ideal gas.
\label{fig:ijxsi}
}
\end{figure}

Although moving beyond the immediate topic of the present paper it is
of interest to consider the behaviour in the deeper parts of the 
convection zone, where the mass can no longer be considered to be constant.
We write $m = M - \Delta m$, where, using equation (\ref{eq:rhosol}),
\begin{equation}
\Delta m = 4 \pi \int_r^R \rho r'^2 \dd r' 
\simeq {4 \pi \Rs^3 \over n+1} \rho_0 \xi^{n+1} \CI_n(\xi) \; ;
\label{eq:msol}
\end{equation}
here, following \citet{Christ1992}, we introduced
\begin{equation}
\CI_n(\xi) = (n+1) \int_0^1 v^n (1 + \xi v)^{-4} \dd v \; ,
\label{eq:curli}
\end{equation}
$v$ being an integration variable,
%\note [Saskia requested a definition of $v$; I am not completely sure
%that it is needed]
defined such that $\CI_n(0) = 1$.
Using this approximation to $m$ in equation (\ref{eq:hydequ}) we obtain
\begin{equation}
u \simeq {\gamma -1 \over \gamma} {G M \over \Rs} \xi 
\left[ 1 - q_0 {\xi^{n+1} \over n+2} \CJ_n(\xi) \right] \; ,
\label{eq:usoli}
\end{equation}
where
\begin{equation}
q_0 = {4 \pi \Rs^3 \rho_0 \over (n+1) M} 
\label{eq:qzero}
\end{equation}
and
\begin{equation}
\CJ_n(\xi) = (n + 2) \int_0^1 v^{n+1} \CI_n(v \xi) \dd v \; ,
\label{eq:curlj}
\end{equation}
which, like $\CI_n$, is defined such that $\CJ_n(0) = 1$.
Plots of $\CI_n$ and $\CJ_n$, for $n = 3/2$, are shown in Fig. \ref{fig:ijxsi}.
Similar expressions are obviously obtained for $c^2$ and $T$.
Also, from equation (\ref{eq:adiab}) we obtain
\begin{equation}
\rho \simeq \rho_0 \xi^n
\left[ 1 - q_0 {\xi^{n+1} \over n+2} \CJ_n(\xi) \right]^n \; .
\label{eq:rhosoli}
\end{equation}

It follows from equation (\ref{eq:usoli}) that $u$, and hence $c^2$ and $T$, 
depend on the adiabatic constant through the term in $q_0$ within the brackets.
This is reflected in the behaviour of $\delta \ln T$ 
in Fig. \ref{fig:moddifrho}.
To make a more quantitative comparison we estimated $\rho_0$ in the two
models by applying equation (\ref{eq:rhozero}) at $r = 0.7 R$ and
determining the corresponding values of $q_0$ from equation (\ref{eq:qzero}),
resulting in $q_0 = 2.07$ and 0.89 for models $\CMrgb$ and $\CMrc$.
Using these values in equation (\ref{eq:usoli}) results in the estimate
of $\delta \ln T$ shown by the thin dot-dashed curve 
in Fig. \ref{fig:moddifrho}.
It is clear that our, relatively rough, approximation captures most of the
effect on the temperature difference.
Similarly, the variation in $\delta \ln \rho$ in the deeper parts of the
convective envelope can be accounted for by equation (\ref{eq:rhosoli}).
It is obvious that the analysis could be iterated by repeating the 
determination of $\Delta m$ using equation (\ref{eq:rhosoli});
little further insight would be obtained from this, however.

The relations for the sound speed are closely 
related to the differences between RGB and clump stars,
noted by \citet{Miglio2012}, in the scaling relation for $\Delta \nu$.
The integral in equation (\ref{eq:asympsep}) is dominated by the contribution
from the convective envelope.
Using the approximation in equation (\ref{eq:csol}) clearly implies the commonly
used scaling relation, i.e., that $\Delta \nu \propto (M/R^3)^{1/2}$.
The correction factor in equation (\ref{eq:usoli}) leads to a decrease in $c$,
an increase in the integral and hence a decrease in $\Delta \nu$, relative
to the simple scaling.
This effect is stronger in the RGB model, owing to the larger value of $q_0$,
than in the clump model, leading to a smaller $\Delta \nu$ in the former
model at fixed mean density, as found by \citet{Miglio2012}.
A more quantitative analysis of this effect and its influence on
the diagnostics based on global asteroseismic parameters would be interesting,
but is beyond the scope of the present paper.

\begin{figure}
\centering
\includegraphics[width=\hsize]{\figdir/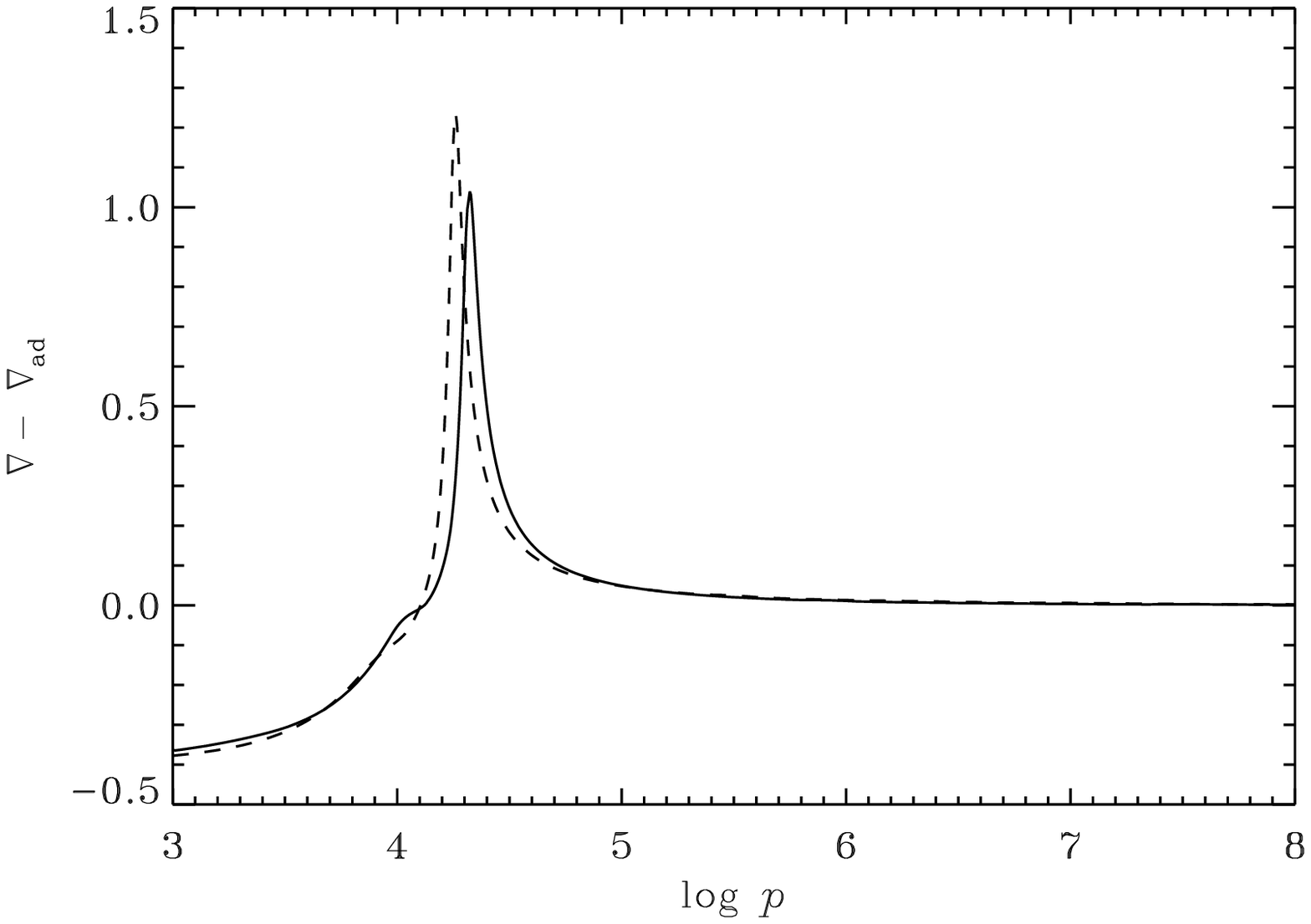}
\caption{Superadiabatic gradient as a function of the logarithm to base 10
of pressure for models $\CMrgb$ (solid line) and $\CMrc$ (dashed line).
\label{fig:superad}
}
\end{figure}

To relate the properties of the convective envelopes to the surface properties
of the stars we note that the bulk of the convection zone is at an essentially
constant specific entropy $s_{\rm ad}$, related to the adiabatic constant
in the region of nearly constant $\Gamma_1$ by
\begin{equation}
s_{\rm ad} \simeq {1 \over \gamma} \ln K \; ,
\label{eq:sad}
\end{equation}
apart from an arbitrary additive constant.
The change in entropy between the photospheric value $s_{\rm ph}$ 
and $s_{\rm ad}$ is determined by the superadiabatic gradient near 
the top of the convection zone,
\begin{equation}
s_{\rm ad} = s_{\rm ph} + \int_{\ln p_{\rm ph}}^{\ln p^*} 
c_p (\nabla - \nabla_{\rm ad}) \dd \ln p \; ,
\label{eq:sad1}
\end{equation}
where $\nabla = \dd \ln T/\dd \ln p$, $\nabla_{\rm ad}$ is its adiabatic
value, $c_p$ is the specific heat at constant pressure,
$p_{\rm ph}$ is the photospheric pressure and the upper limit 
of the integral is at a
point in the adiabatically stratified interior of the convection zone.
The superadiabatic gradients in models $\CMrgb$ and $\CMrc$ are illustrated
in Fig. \ref{fig:superad}.
It is evident that $\nabla - \nabla_{\rm ad}$, and hence the integral in
equation (\ref{eq:sad1}), is bigger in $\CMrc$ than in $\CMrgb$;
this dominates the fact that $K$ is bigger in the former model,
corresponding to the lower density and hence the difference in the ionization
of helium leading to the difference in $\epsilonc$.
The behaviour of $\nabla$ in the superadiabatic region, at the assumed fixed
mixing length, is dominated by the fact that the opacity is higher in
model $\CMrc$ than in $\CMrgb$, owing to the higher effective temperature
and hence higher photospheric temperature and the strong temperature dependence
of the dominant ${\rm H}^-$ opacity.

This leaves open the question of the dominant causation in the difference
between the thermodynamic state of the two models.
It seems plausible to us that the main effect must be the higher mass in the
core and the consequent lower density in the convective envelope, at fixed
stellar radius.
The properties of the superficial parts of the convective envelope have
to respond to this, requiring the higher superadiabatic gradient and hence
the higher effective temperature.
Further investigations of this interplay between the interior and superficial
properties of red giants appear worthwhile.

\label{lastpage}

\end{document}